\documentclass[aps,prl,twocolumn,showpacs,10pt,superscriptaddress,floatfix,longbibliography]{revtex4-2}

\usepackage{newtxtext}
\usepackage{newtxmath}
\usepackage{graphicx}
\usepackage{subfigure}
\usepackage{amsmath}
\usepackage{amsfonts}
\usepackage{mathrsfs}
\usepackage{bbm}
\usepackage{subfigure}
\usepackage{booktabs}
\usepackage{xcolor}
\usepackage[colorlinks=true]{hyperref}
\usepackage{ulem}
\usepackage[T1]{fontenc}
\usepackage{svg}
\usepackage{titlesec}
\usepackage{lipsum}  

\newcounter{appsec}
\renewcommand{\theappsec}{Appendix~\Alph{appsec}}

\titleformat{\section}
  [runin]                        
  {\normalfont\itshape\color{blue}} 
  {}                             
  {0.5em}                          
  { \theappsec : }                             
  [{\color{black}--}]                            

\titlespacing*{\section}{0.5em}{0pt}{0pt} 

\let\oldsection\section
\renewcommand{\section}{\refstepcounter{appsec}\oldsection}

\newcommand{\blue}[1]{\textcolor{blue}{#1}}

\begin{document}

\title{Berry Curvature Induced Spin Nernst and Thermal Edelstein Effects in Proximity Superconductors}

\author{Zhen-Cheng Liao}
\affiliation{School of Physics, Sun Yat-sen University, Guangzhou 510275, China}

\author{Cong Xiao}
\email{congxiao@fudan.edu.cn}
\affiliation{Interdisciplinary Center for Theoretical Physics and Information Sciences (ICTPIS), Fudan University, Shanghai 200433, China}

\author{Zhi Wang}
\email{wangzh356@mail.sysu.edu.cn}
\affiliation{School of Physics, Sun Yat-sen University, Guangzhou 510275, China}
\affiliation{Guangdong Provincial Key Laboratory of Magnetoelectric Physics and Devices, Sun Yat-sen University, Guangzhou 510275, China}

\author{Qian Niu}
\affiliation{School of Physics, University of Science and Technology of China, Hefei, Anhui 230026, China }

\begin{abstract}
We propose two thermo-spintronic responses in proximity induced superconductors with spin-orbit coupled band - spin Nernst effect and thermal Edelstein effect stemming respectively from momentum-space and mixed superconducting Berry curvatures. We unveil that the Bloch band spin-orbit coupling and pairing are entangled in shaping the superconducting Berry curvatures rather than in an additive manner. The resulting thermo-spintronic responses are significantly enhanced at low temperatures by the emergence of Bogoliubov Fermi surface, and can be effectively controlled by tuning the latter. The effects render a probe of Bogoliubov Fermi surface and superconducting Berry curvatures supported by it. Our work reveals Berry curvature properties in topologically trivial proximity superconductors with spin-orbit coupling, and opens a new route to superconducting spintronics.
\end{abstract}
\maketitle

Spin-orbit coupling (SOC) of Bloch electrons is an important origin of Berry curvature \cite{xiao2010berry}, which characterizes the local quantum geometry in the parameter space of the electronic wave function and underlies a variety of fundamental physical effects \cite{jungwirth2002anomalous,fang2003anomalous,yao2004first,murakami2003dissipationless,xiao2006berryphase,xiao2012coupled,yang2014dirac}. In superconductors, SOC may profoundly influence the properties of superconducting ground state and quasiparticles. For intrinsic superconducting systems such as noncentrosymmetric superconductors, the presence of SOC mixes the spin-singlet and spin-triplet Cooper pairing, inducing a variety of nontrivial properties \cite{gorkov2001superconducting,smidman2017superconductivity,fischer2023}. For proximity induced superconductivity \cite{amundsen2024}, the conventional $s$-wave pairing can induce effective chiral superconducting gap for electrons on a particular SOC band \cite{fu2008superconducting,zhang2008p+ip,sato2009nonabelian,sau2010generic,Alicea2010,mao2011superconducting,Potter2011,Nagaosa2012Bilayer,liu2013d+id,shi2020attractive,qi2011topological,sato2017topological}, which has a topological origin in the SOC-induced Berry phase accumulated on the electron Fermi surface \cite{zhang2008p+ip,mao2011superconducting,shi2020attractive}.

While the topological aspect of proximity induced chiral superconductivity with SOC  \cite{frolov2020topological,flensberg2021engineered}, i.e., Chern number \cite{read2000paired,schnyder2008classification} and Berry phase \cite{qin2019chiral}, has been made clear, the local Berry curvature properties have largely been unexplored \cite{liang2017wavepacket,Hsu2025}, and their possible roles in nonequilibrium spintronic phenomena of proximity superconductors remain unclear. Filling in this gap will not only uncover fundamental physics of SOC-shaped Berry curvature in superconductors but also reveal new physical effects in superconducting spintronics \cite{Linder2015superconducting,Han2020spin,Yang2021boosting}.

In this letter, we focus on proximity-induced superconductors with SOC and propose two thermo-spintronic responses: the spin Nernst effect induced by $k$-space Berry curvature of superconductors, and the thermal Edelstein effect induced by a mixed Berry curvature. The spin Hall transport and Edelstein effect are two cornerstone effects of spintronics \cite{manchon2019}, and when the temperature gradient serves as the driving force, it is the spin Nernst effect \cite{Bose2018SNE,Bose2018SNT,kim2020SNT,zhang2020SNE,jain2023thermally} and the thermal Edelstein effect \cite{freimuth2014DMI-SOT,xiao2016spin,Shitade2019,dong2020berry} that come into play. To the best of our knowledge, these two thermo-spintronic responses of origins in superconducting Berry curvatures have not been studied before.

To reveal the underlying physics, we derive general forms of the two types of superconducting Berry curvatures near the electron Fermi surface, and unveil non-additive entanglement of pairing and SOC single-particle physics in shaping the superconducting Berry curvature. Motivated by recent experimental progress in generating Bogoliubov Fermi surface in proximity-induced superconductors \cite{zhu2021discovery,Phan2022BFS}, we consider the Berry spintronic responses upon the emergence of Bogoliubov Fermi surface induced by Doppler shift in a ferromagnetic Rashba model in proximity to a conventional $s$-wave superconductor. We show that the responses are significantly enhanced at low temperatures and qualitatively change the behaviors after the Bogoliubov Fermi surface is created. These responses thus provide a direct probe of Bogoliubov Fermi surface and superconducting Berry curvatures on it.
Our findings may motivate fundamental studies on superconducting quantum geometry in proximitized electronic systems and open up the new research direction of superconducting thermo-spintronics.

{\it \blue {Formulation of response}}--
Proximity-induced superconductivity can be well described by a mean-field Bogoliubov-de Gennes (BdG) Hamiltonian. In the Nambu basis for the Bloch electron operators, the BdG matrix can be written in the band representation as \cite{supp}
\begin{equation}\label{eq:BdG}
{H}_{\bf k}=\left(\begin{array}{cc}
{h}_{nn'}\left(\mathbf{k}\right) & \tilde{\Delta}_{nn'}\left(\mathbf{k}\right)\\
\tilde{\Delta}_{nn'}^{\dagger}\left(\mathbf{k}\right) & -{h}_{nn'}^{*}\left(-\mathbf{k}\right)
\end{array}\right).
\end{equation}
Here ${h}_{nn'\mathbf{k}}=\xi_{n{\bf k}} \delta_{n,n'}$ is the electronic spectra, $\tilde{\Delta}_{nn'}\left(\mathbf{k}\right)$ is the effective gap function in the band representation:
\begin{equation}\label{eq:gapfunction}
  \tilde{\Delta}_{nn'\mathbf{k}}= \Delta_{0} \int_{{\bf R},{\bf r}} \phi_{n\mathbf{k},\sigma }^{*} ({\bf R}+\frac{\bf r}{2}) {\chi}^c_{\sigma\sigma'}({\bf r}) \phi_{n'-\mathbf{k},\sigma'}^{*}({\bf R}-\frac{\bf r}{2}),
\end{equation}
which is obtained via the Fourier transformation of a $s$-wave proximity induced gap with the electronic Bloch function. Here, $\Delta_0$ is the amplitude of the s-wave proximity-induced gap, and $\chi^c$ is the cell-periodic part of the gap. ${\bf R}$ and ${\bf r}$ are the center-of-mass and relative coordinates of the two electrons in a Cooper pair, $\phi_{n \mathbf{k}, \sigma}$ is the $\sigma$ component of the electronic spinor cell-periodic Bloch function with band $n$ and momentum $\hbar \bf k$, and $\int_{{\bf R},{\bf r}}$ is the unit-cell integral. The Einstein summation convention is taken for $\sigma$. 
In the present context, we assume the spinor structure of $\phi_{n \mathbf{k}}$ is caused solely by SOC.
With the Hamiltonian (\ref{eq:BdG}), we can construct Berry curvatures in pertinent parameter spaces of BdG wave functions and derive the Berry curvatures induced spin Nernst effect and thermal Edelstein effect. 

The spin Nernst effect describes the generation of a spin current transverse to the applied temperature gradient.
Here, we adopt the definition of bulk conserved spin current \cite{shi2006proper,Murakami2006conserved,xiao2021conserved}, which certifies two plausible physical properties: (1) The equilibrium spin current is at most a circulating magnetization current that cannot flow out of the sample; (2) After subtracting this circulating magnetization current from the total spin current, the obtained transport spin current driven by temperature gradient is indeed the Onsager reciprocal of thermal current driven by Zeeman gradient (gradient of a Zeeman field \cite{Fisher1999}) \cite{xiao2021conserved,Xiao2021conserved-arxiv}. In \ref{app:A}, we present the theory of this conserved spin current, whereas the result for spin Nernst effect can be reached intuitively as follows. The entropy current density ${\bf j}^{\mathfrak{s}}$ (which is proportional to the thermal current density ${\bf j}^Q$ as ${\bf j}^{\mathfrak{s}}={\bf j}^Q/T$, with $T$ as the temperature) driven by Zeeman gradient is given by   
\begin{equation}
{\bf j}^{\mathfrak{s}} = \int_{\mathbf{k}} (-\dot {\bf k} \times \Omega_{\mathbf{k}})\mathfrak{s},
\label{inverse SNE}
\end{equation}
where $-\dot {\bf k} \times \Omega_{\mathbf{k}}$ is the anomalous velocity of the superconducting quasiparticle, with $\dot {\bf k}=-s_\alpha {\bf \nabla}m_\alpha$ being the Zeeman-gradient force \cite{xiao2020unified,pan2024thermo}.
Here, $\Omega_{\bf k}$ and $s_{\alpha}$ are the $k$-space Berry curvature \cite{Gradhand2014berry} and the spin expectation value of the quasiparticle, respectively. The summation is implied over repeated Cartesian indices ($\alpha, \beta$, ...). $\mathfrak{s}=-\partial g/\partial T$ is the state resolved entropy density, with $g = k_{\text{B}} T \ln(1 - f)$ being the grand potential and $f$ as the Fermi distribution. Equation (\ref{inverse SNE}) is nothing but the inverse spin Nernst effect, and according to the Onsager reciprocity relation, the spin Nernst effect is described by the Hall spin current
\begin{equation}
\mathcal{J}^{s_{\alpha}} = \mathbf{\alpha}^{\alpha}_H \times (-\nabla T),
\end{equation}
with the spin Nernst conductivity given by \cite{xiao2021conserved}
\begin{equation}
\mathbf{\alpha}^{\alpha}_H = \int_{\mathbf{k}} \mathfrak{s} \Omega_{\mathbf{k}}s_{\alpha}.
\label{eq:ASN}
\end{equation}

The thermal Edelstein effect describes the spin generation by a temperature gradient
\begin{equation}
\delta S_\beta = \chi_{\alpha \beta}(-\nabla_\alpha T),
\end{equation}
where we focus on the thermal Edelstein coefficient $\chi_{\alpha \beta}$ determined solely by the band structure of superconducting quasiparticles. 
The direct derivation is given in \cite{supp}, and here we again utilize the Onsager reciprocity to get the result readily. The Onsager reciprocal of the thermal Edelstein effect is the thermal current pumped by the time variation rate of Zeeman field $\dot{\bf m}$ \cite{freimuth2016inverse}. According to the semiclassical equation of motion in superconductors \cite{sundaram1999wavepacket,wang2021berry}, the latter effect is formulated as
\begin{equation}
{\bf j}^{\mathfrak{s}} = -\int_{\mathbf{k}} \Omega_{\mathbf{k}t}\mathfrak{s},
\label{inverse TE}
\end{equation}
where the momentum-time (${\bf k},t$) space Berry curvature $\Omega_{\mathbf{k}t}=\Omega_{\mathbf{km}}\dot{\bf m}$ is the (minus) anomalous velocity induced by $\dot{\bf m}$, which encodes the mixed Berry curvature tensor $\Omega_{\mathbf{km}}$ defined in the phase space spanned by $\bf k$ and $\bf m$. (In the normal state, $\Omega_{\mathbf{km}}$ underlies a variety of magnetoelectric coupling phenomena \cite{Franz2010,freimuth2014,Freimuth2015Onsager,yuriy2017,xiong2018,Yuriy2019mixed,xiao2021sot,xiao2023,tang2024lossless,Manchon2024pumping}.)
The Onsager reciprocity then tells us that
\begin{equation}\label{eq:TE}
\chi_{\alpha 
\beta} = \int_{\mathbf{k}} \mathfrak{s} \Omega_{k_\alpha m_\beta}|_{{\bf m} \rightarrow 0}.
\end{equation}
Here $\mathbf{m}$ is an auxiliary Zeeman field that is taken to be zero at the last step of the calculation \cite{dong2020berry,xiao2022intrinsic}. 

The thermal Edelstein effect is distinct from the previously discussed Edelstein effect in superconductors \cite{Edelstein1995SC,raines2019manifestations,ando2024spin,He2021SCOME,Hu2025NLSCME}. The latter is a response to supercurrent at equilibrium, while the former is a nonequilibrium response. They are different physical effects and have different symmetry requirements. The equilibrium Edelstein effect requires breaking inversion symmetry ($\mathcal{P}$) but can appear in nonmagnetic systems. In contrast, the thermal Edelstein effect induced by superconducting Berry curvature requires breaking both $\mathcal{P}$ and time reversal ($\mathcal{T}$), because spin and temperature gradient have opposite transformation properties under both $\mathcal{P}$ and $\mathcal{T}$.

{\it \blue {Superconducting Berry curvatures}}--
Superconducting pairing that appears near the electron Fermi surface is most relevant to nonequilibrium effects. This is the intraband component $\tilde{\Delta}_{nn}$ of the effective gap function (\ref{eq:gapfunction}). In the following, we analyze the superconducting Berry curvature around such gaps, where the BdG equation reduces to 
\begin{eqnarray}
\left(\begin{array}{cc}
\xi_{n,\mathbf{k} } &\tilde{\Delta}_{nn\mathbf{k}} \\
\tilde{\Delta}^{*}_{nn\mathbf{k} }  & -\xi_{n,-\mathbf{k}} 
\end{array}\right)
\left(\begin{array}{cc}
\mu \\
\nu
\end{array}\right)
= E \left(\begin{array}{cc}
\mu \\
\nu
\end{array}\right).
\end{eqnarray}
Here $\mu$ and $\nu$ are Bogoliubov amplitudes associated with the BdG eigenstate $|\psi_{{n\bf k}}\rangle  = [\mu |\phi_{n{\bf k}}\rangle, {\nu} |\phi^*_{n-{\bf k}}\rangle]^{\mathrm{T}}$. The $k$-space Berry connection $ \mathbf{\mathcal{A}}=\langle\psi_{n{\bf k}}|i\nabla_{\bf k}\psi_{n{\bf k}}\rangle$  derived from this eigenstate reads \cite{supp}
$\mathbf{\mathcal{A}}({\bf k})
=-\frac{1}{2}\rho \nabla_{{\bf k}}\varphi +  |\mu|^2 {\bf \mathcal{A}}_{n}^{b}({\bf k}) +|\nu|^2{\bf \mathcal{A}}_{n}^{b}(-{\bf k})$, where $\rho = |\mu|^2 - |\nu|^2$ is the effective charge of the quasiparticle, $\varphi =\arg(\tilde{\Delta}_{nn})$, and ${\bf \mathcal{A}}^{b}_{n}=\langle\phi_{n{\bf k}}|i\nabla_{\bf k}\phi_{n{\bf k}}\rangle$ is the electronic Berry connection.
The $k$-space Berry curvature $\mathbf{\Omega}_{\bf k}=\nabla_{{\bf k}}\times \mathbf{\mathcal{A}}({\bf k})$ is thus given by
\begin{eqnarray}\label{eq:BCFS}
\mathbf{\Omega}_{\bf k}
=\frac{1}{2}D_{\bf k}\varphi \times \nabla_{{\bf k}}\rho +  |\mu|^2 {\bf \Omega}^{b}_{n}({\bf k}) -|\nu|^2{\bf \Omega}^{b}_{n}(-{\bf k}),
\end{eqnarray}
where ${\bf \Omega}_{n}^{b}=\nabla_{{\bf k}}\times \mathbf{\mathcal{A}}_{n}^{b}$ is the electronic Berry curvature, and 
\begin{eqnarray}\label{eq:k-derivative}
    D_{\bf k}\varphi = \nabla_{{\bf k}}\varphi - \mathbf{\mathcal{A}}_{n}^{b}({\bf k}) + {\bf \mathcal{A}}_{n}^{b}({\bf -k})
\end{eqnarray} is a gauge invariant $k$-derivative of the superconducting phase. It is interesting to note that $D_{\bf k}\varphi $ is the $k$-space counterpart of supercurrent velocity $D\varphi={\bf \nabla\varphi}-2{\bf A}$, where $\bf A$ is the magnetic vector potential. Its $k$-space U(1) gauge invariance manifests the self-consistency between order parameter and quasiparticle of the induced effective superconductivity \cite{supp}.

\begin{figure*}[htb]
    \centering
    \includegraphics[width=1\linewidth]{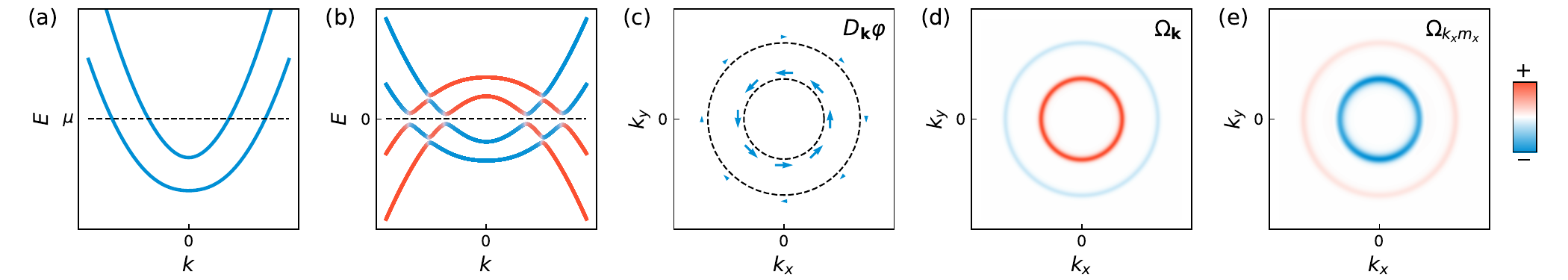}
    \caption{(a) Spin-split electron spectrum of the ferromagnetic Rashba model. The chemical potential cuts both bands. (b) The corresponding BdG spectrum of the model in proximity with a $s$-wave superconductor, where color represents the effective charge of the quasiparticle. (c) Gauge invariant $k$-space derivative of the superconducting phase $D_{\bf k}\varphi$, (d) $k$-space Berry curvature $\Omega_{\bf k}$, and (e) phase-space Berry curvature $\Omega_{k_x m_x}$ in the two small-gap regions of the BdG spectrum around zero energy. The length of arrows in (c) signifies the strength of $D_{\bf k}\varphi$. }
    \label{fig:Figure1}
\end{figure*}

All the terms in Eq. (\ref{eq:BCFS}) are connected to superconductivity, and manifest two prominent features: (1) the entanglement of superconductivity pairing and electronic wave functions in shaping the superconducting Berry curvature, and (2) the co-presence of electronic Berry quantities at both $\bf k$ and $\bf -k$. These two features are rooted in the nature of superconducting quasiparticle being a superposition of electron at ${\bf k}$ and hole at $-{\bf k}$. However, they are hidden in the presence of $\mathcal{T}$ symmetry of electronic Hamiltonian (details in \ref{app:B}), where the Berry quantities are simply additive terms of pairing and single-particle contributions. The developed theory therefore unveils a profound difference in $k$-space quantum geometry of superconductivity for $\mathcal{T}$-invariant and $\mathcal{T}$-breaking electronic systems. 



The analysis on the mixed Berry curvature $\Omega_{\mathbf{km}}$ is similar, and is presented in \ref{app:C}. One also finds the mentioned two characteristics of superconducting $\Omega_{\mathbf{k}}$. In particular, the pairing and single-electron band physics contribute non-additively to the superconducting $\Omega_{\mathbf{km}}$.

{\it \blue {Berry curvature and thermo-spintronic responses in topologically trivial proximity superconductors}}--We illustrate the above general reasoning by a ferromagnetic Rashba model in proximity contact with a conventional $s$-wave superconductor \cite{sau2010generic}. This model captures the basic characters of superconductor/ferromagnet heterostructures, and is suitable for illustrating the proposed effects. The single-particle Hamiltonian reads
\begin{equation}\label{eq:modelHamiltonian}
    \mathcal{H}_0=  h_0 +  \mathbf{h} \cdot\boldsymbol{\sigma},
\end{equation}
where $h_0 = k^2/2m - \mu$ with $m$ the mass of the electron and $\mu$ the chemical potential, $\boldsymbol{\sigma}$ is the vector Pauli matrix for spin, and ${\bf h} = (\alpha_R k_y, -\alpha_R k_x, V_z)$ with $\alpha_R$ the Rashba SOC coefficient and $V_z$ the Zeeman energy. As shown in Fig.~\ref{fig:Figure1}(a), the two bands are fully split by the Zeeman energy. We choose a Fermi level cutting both bands, yielding two electron Fermi surfaces. The proximity effect induces a $s$-wave superconducting gap $\Delta$ \cite{zhang2008p+ip,sato2009nonabelian,sau2010generic}. The resulting BdG spectrum is shown in Fig.~\ref{fig:Figure1}(b), with two small gaps at the conduction band minimum corresponding to the two electron Fermi surfaces. Around these small gap regions, the intra-band effective gap function from Eq.~(\ref{eq:gapfunction}) is acquired as \cite{supp,Alicea2010} 
\begin{equation}
   \tilde{\Delta}_{\bf k}=\Delta \sin \theta_{\bf k} e^{\pm i \zeta_{\bf k}}
   ={\alpha_R\Delta(k_y \pm ik_x)}/{\sqrt{\alpha_{R}^2 k^2+V^2_z}},  
\end{equation}
where $\theta_{\bf k}$ and $\zeta_{\bf k}$ are respectively the polar and azimuthal angles of the vector ${\bf h}$, and $\pm$ represents the two electronic bands. $\tilde{\Delta}_{\bf k}$ has a chiral $p$-wave symmetry.
The gauge invariant phase derivative $ D_{\bf k}\varphi_{\bf k}$ thus exhibits a chiral pattern with contrasted chirality around the two gap regions (Fig.~\ref{fig:Figure1}(c)).

The $k$-space superconducting Berry curvature around the two gap regions is given by (hereafter $\Omega_{\mathbf{k}}$ in 2D systems is denoted as a scalar)
\begin{equation} \label{eq:BC2L}
\Omega_{\mathbf{k}}
    =   \pm\frac{1}{2}\cos\theta_{\bf k}(\nabla_{\mathbf{k}}\zeta_{\bf k}\times \nabla_{\mathbf{k}}\rho)_z + \rho \Omega^b_{\bf k},
\end{equation}
with $\Omega_{\mathbf{k}}^{b} = \pm \mathbf{h}\cdot (\partial_{k_x} \mathbf{h} \times \partial_{k_y} \mathbf{h})/|\mathbf{h}|^3$. As $\rho \approx 0$ near the gap regions, the first term in Eq.~(\ref{eq:BC2L}) dominates (Fig.~\ref{fig:Figure1}(d)).
This term resembles the Berry curvature of intrinsic chiral superconductivity \cite{wang2021berry,supp}, but with a difference of $\cos \theta_{\bf k}$. Thus, even if the effective pairing $\tilde{\Delta}_{\mathbf{k}}$ takes a chiral form, the superconducting Berry curvature is distinct from that of an intrinsic chiral superconductor. A particular example is the interface between a topological insulator and a conventional superconductor \cite{fu2008superconducting} (${\bf h} = \alpha_R (k_x, k_y, 0)$), where $\tilde{\Delta}_{\mathbf{k}} = \Delta( k_x+ik_y)/k$ is of chiral $p$-wave form but $\Omega_{\mathbf{k}}=0$.


We then discuss the mixed Berry curvature. The in-plane components of $\Omega_{k_\alpha m_\beta}$ are physically relevant because the corresponding induced spin polarizations are transverse to the magnetization hence can exert a torque on the latter. Around the gap regions, we have $D_{\bf m}\varphi = 0$ and $\nabla_{\bf m} \rho =0$ \cite{supp}, thus, according to Table \ref{tab:km} in \ref{app:C} and the $C_{2z}$ symmetry of the model, one gets $\Omega_{k_\alpha m_\beta}= \Omega_{k_\alpha m_\beta}^{b}=\Omega_{k_\alpha h_\beta}^{b}$. This result is remarkable as it means $\Omega_{\bf k m}$ of the superconductor is given solely by the normal state. Because of this connection, one further has $\Omega_{{ k}_x {m}_y}=0$, and
$\Omega_{{ k}_x {m}_x}=\Omega^b_{\bf k}/\alpha_R$ (Fig.~\ref{fig:Figure1}(e)).



\begin{figure}[t]
    \centering
    \includegraphics[width=1.0\linewidth]{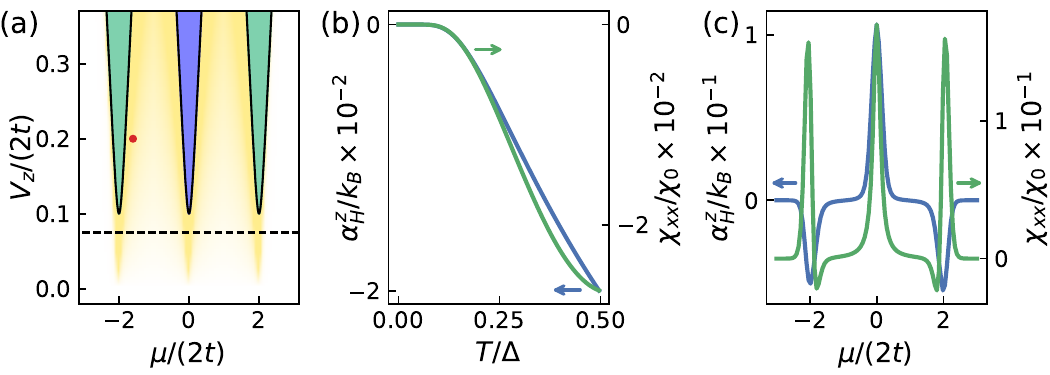}
    \caption{(a) Phase diagram of the tight-binding proximitized ferromagnetic Rashba model. Topological phases possess Chern number $C = -1$ (green) and $C = 2$ (purple). The yellow color denotes the maximum $k$-space Berry curvature in topologically trivial region. (b) Superconducting spin Nernst conductivity $\alpha_H^z$ and thermal Edelstein conductivity $\chi_{xx}$ as a function of temperature, calculated with the parameters of the red point in (a), $\alpha_{R}/2t=0.4$ and $\Delta/2t=0.1$. Here, $\chi_0=\hbar k_B/ 2ta$ with $a$ the lattice constant and $t$ the nearest-neighbor hopping. (c) $\chi_{xx}$ and $\sigma_H^z$ at $T/\Delta=0.5$ as a function of electron chemical potential along the dashed line.}
    \label{fig:FigureC}
\end{figure}

Next, we calculate the thermo-spintronic responses (\ref{eq:ASN}) and (\ref{eq:TE}).
For quantitative evaluations, we take the tight-binding version of the above proximitized Rashba model (\ref{app:tight-binding}). The topological phase diagram is shown in Fig.~\ref{fig:FigureC}(a). It is noted that the Berry curvatures can be prominent in topologically trivial regions, inducing thermal spintronic responses as shown in Fig.~\ref{fig:FigureC}(b). Because of the presence of global excitation gap, both signals are nonzero only due to quasiparticle excitations at finite temperatures. In Fig.~\ref{fig:FigureC}(c), we plot the responses as a function of electron chemical potential, which exhibit peaks at hot spots of Berry curvatures in the BdG spectrum.

\begin{figure}[t]
    \centering
    \includegraphics[width=1\linewidth]{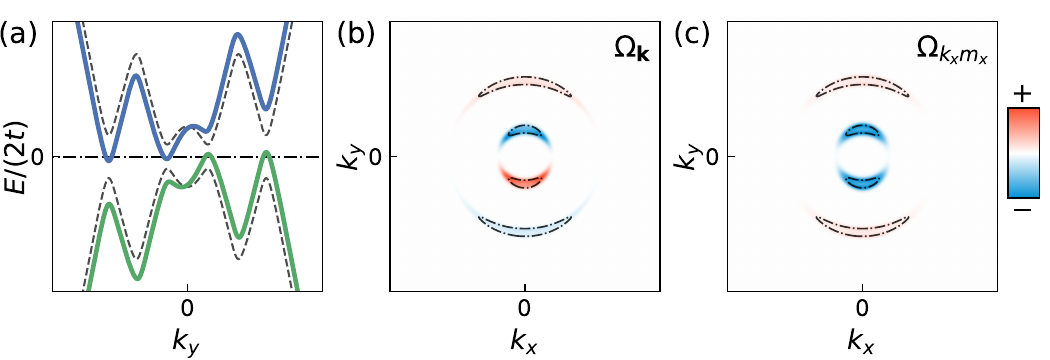}
    \caption{Bogoliubov Fermi surface created by Doppler shift in the tight-binding ferromagnetic Rashba model proximity to a $s$-wave superconductivity. (a) The BdG spectrum before (dashed line) and after (solid line) considering the Doppler shift induced by a supercurrent ${\bf q}=0.12\pi\hbar/a\hat{\bf y}$. (b) The $k$-space Berry curvature and (c) $xx$ component of the mixed Berry curvature on the Bogoliubov Fermi surface. }
    \label{fig:Figure2}
\end{figure}

{\it \blue {Bogoliubov Fermi surface}}--Compared to fully gapped superconductors, thermo-spintronic responses at low temperatures can be strongly enhanced upon the emergence of Bogoliubov Fermi surface. This is because the entropy density $\mathfrak{s}$, which appears in both response functions, is symmetrically peaked around zero energy and reduces to \cite{xiao2020unified} $\mathfrak{s}=\frac{1}{3}\pi^2 k_B^2 T \delta (E_{\bf k})$ at low temperatures, with $E_{\bf k}$ as the quasiparticle energy. As such, $\mathfrak{s}$ is peaked on the Bogoliubov Fermi surface, and decreases dramatically away from it. The low-temperature responses thus render a way to probe Berry curvatures ${ \Omega}_{ \bf k}$ and $\Omega_{\bf k m}$ on the Bogoliubov Fermi surface.

\begin{figure}[t]
    \centering
    \includegraphics[width=1.0\linewidth]{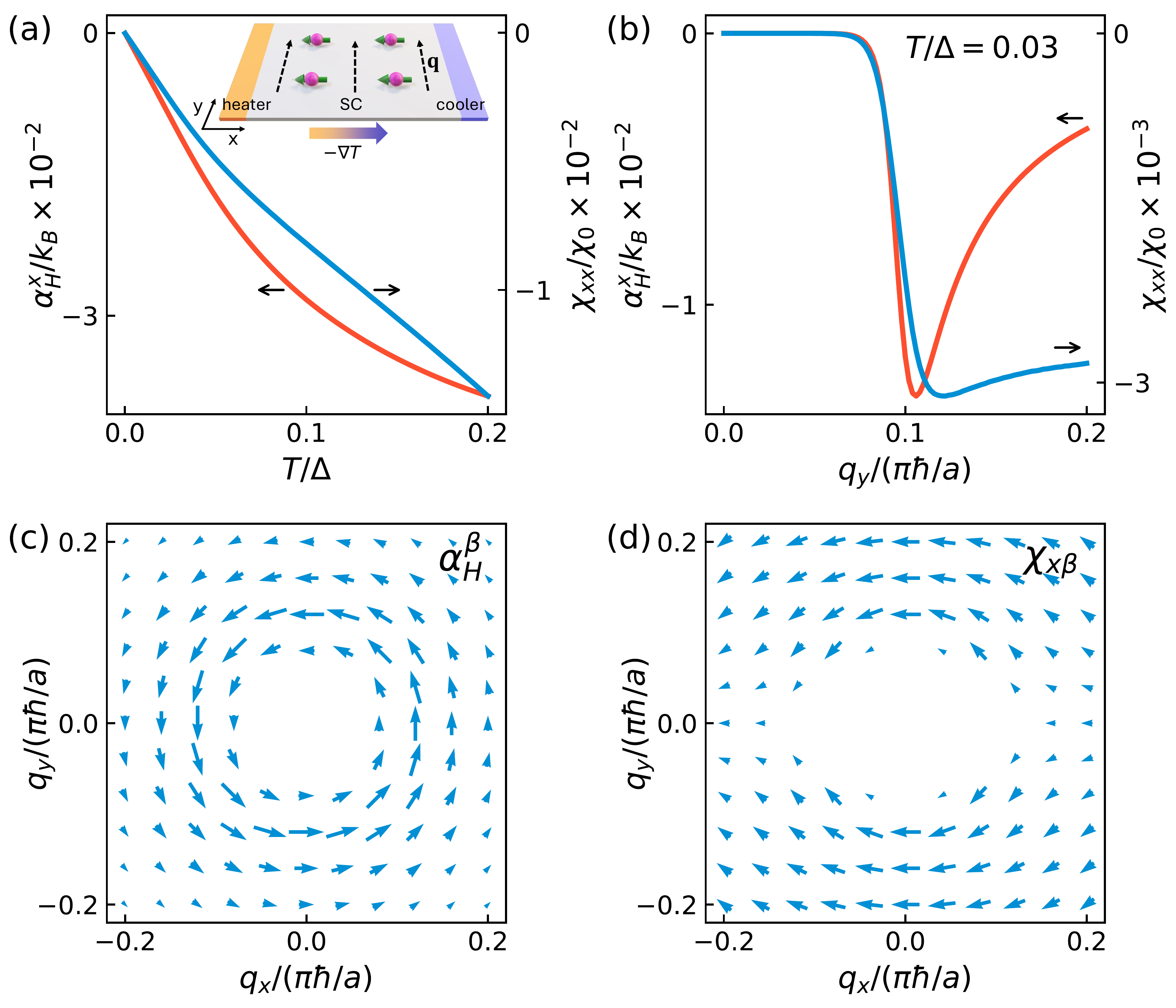}
    \caption{Superconducting thermo-spintronic responses of Bogoliubov Fermi surface. (a) Temperature dependence of $\alpha_H^x$ and $\chi_{xx}$ for the Bogoliubov Fermi surface of Fig. \ref{fig:Figure2}. The inset illustrates the spatial relations of the temperature gradient, supercurrent and induced spin in the thermal Edelstein effect.} (b) $\alpha_H^x$ and $\chi_{xx}$ as a function of supercurrent magnitude at a fixed temperature. 2D vector plot of the supercurrent dependence of (e) $\alpha_H^\beta$ and (f) $\chi_{x\beta}$. Here, $\alpha_H^\beta$ and $\chi_{x\beta}$ are plotted as vectors with respect to the Cartesian index $\beta=x,y$. The orientation and length of arrows signify the direction and magnitude of the vectors, respectively. Model parameters are taken the same as Fig.~\ref{fig:FigureC}(a).
    \label{fig:Figure3}
\end{figure}

The Bogoliubov Fermi surface can be induced and tuned by Doppler shift of the BdG spectrum through injected
supercurrent (in-plane external magnetic field) \cite{volovik1993superconductivity,yuan2018zeemaninduced,zhu2021discovery,Phan2022BFS}. When the Doppler shift is large enough, the BdG spectrum will exhibit zero energy areas (Fig.~\ref{fig:Figure2}(a)), i.e., the Bogoliubov Fermi surfaces. We plot the Bogoliubov Fermi surface of the considered model under a supercurrent ${\bf q} =0.12 \pi\hbar/a \hat{\bf y}$, and show the significant distribution of $k$-space Berry curvature and mixed Berry curvature on the Bogoliubov Fermi surface in Fig.~\ref{fig:Figure2}(b) and Fig.~\ref{fig:Figure2}(c). Because of these strong Berry curvatures, the thermal spintronic responses are greatly enhanced at low temperatures, as shown in
Fig.~\ref{fig:Figure3}(a), in sharp contrast to the vanishing result in the absence of Bogoliubov Fermi surface (Fig.~\ref{fig:FigureC}(b)). One also finds that the emergence of Bogoliubov Fermi surface qualitatively alters the temperature dependence of the response signals. A linear relation is observed as a result of entropy density being $\mathfrak{s}\sim T$, serving as a hallmark of Bogoliubov Fermi surface.

The Doppler shift can effectively tune the geometry of the Bogoliubov Fermi surface and quantum geometry on it, thus rendering a unique route to fine-tuning of superconducting spintronic responses.
Figure~\ref{fig:Figure3}(b) displays the tuning of $\alpha_H^x$ and $\chi_{xx}$ by varying the magnitude of supercurrent. They become appreciable only after the supercurrent reaches $q=0.1\pi\hbar/a$, at which the Bogoliubov Fermi surface emerges. More interestingly, one can vary both the magnitude and orientation of the supercurrent in order to fully control the shape of Bogoliubov Fermi surfaces and thus the superconducting quantum geometric responses.
Figure~\ref{fig:Figure3}(c) and (d) demonstrate such a systematic tuning of the spintronic responses by supercurrent vector in 2D $\bf q$ space. In these two figures, $\alpha_H^\beta$ and $\chi_{x\beta}$ are vectors with respect to the Cartesian index $\beta=x,y$. The length of the arrows denotes the magnitude of these two vectors, and the orientation of the arrows indicates the spin polarization direction of the induced spin current (Fig.~\ref{fig:Figure3}(c)) or spin accumulation (Fig.~\ref{fig:Figure3}(d)).

{\it {\color{blue} Discussion.}}--
We have revealed Berry curvature properties induced by the interplay of Bloch band SOC and superconducting pairing in proximity superconductors, and proposed the resulting spin Nernst effect and thermal Edelstein effect. Our calculations on the well known proximitized ferromagnetic Rashba model show that these Berry curvature effects can be significant in the previously less studied topologically trivial superconducting regime.
These two effects lay a groundwork for superconducting spin-caloritronics, which may open a new research area in both fields of superconductivity and spintronics.

For example, on the side of superconductivity study, we have unveiled significantly enhanced superconducting spintronic responses and their characteristic temperature dependence upon the emergence of Bogoliubov Fermi surface, which in turn render a probe of the latter and its Berry curvature properties. In the context of spintronics, the Edelstein effect induced by Rashba spin-orbit coupling in ferromagnet/nonmagnet hyrbid structures can generate a spin-orbit torque in the ferromagnet \cite{miron2010,miron2011}. In parallel, our studied superconducting thermal Edelstein effect in ferromagnet/superconductor hyrbids is anticipated to induce a thermal spin-orbit torque in the ferromagnet. In the proximitized Rashba system, the Berry curvature induced spin polarization $\delta {\bf S}$ is odd in the magnetization, thus the resulting torque is an antidamping like torque, which is much more efficient than field-like torques \cite{manchon2019}.

In addition to the Berry curvature effects considered here, there may also be extrinsic contributions from quasiparticle scattering on impurities to superconducting thermo-spintronic responses. Like in the thermally driven transport in normal state, the Berry curvature effect may dominate over the extrinsic terms in moderately dirty samples 
\cite{Ong2004ANE,Miyasato2007,Shi2008ANE,Yuriy2013ANE,Behnia2017Mn3Sn,Arita2017ANE,Felser2022giant}. On the other hand, in the dirty limit where the interband structure is severely smeared by disorder, there is no Berry curvature effect (which encodes interband coherence \cite{xiao2010berry}), and all the contributions are extrinsic \cite{Manchon2012diffusive,matsushita2022spinnernst}.
Detailed and systematic studies on extrinsic contributions across broad ranges of disorder concentrations are left for future efforts.

\bibliography{citations}


\begin{widetext}
\subsection{    End Matter}
\end{widetext}

\appendix 
\makeatletter
\renewcommand{\theequation}{\Alph{appsec}.\arabic{equation}}
\renewcommand{\thefigure}{\Alph{appsec}\arabic{figure}}
\@addtoreset{equation}{appsec}  
\makeatother

\section{Conserved spin current}\label{app:A}
In the literature on spin transport, the spin current is conventionally defined as the anticommutator of the spin and velocity $\hat{\bf J}^{s}=\{\hat{s},\hat{\bf v}\}/2$ \cite{sinova2015spin}, which can be called the conventional spin current (For simple notations, in this section we use $\hat{s}$ to denote a component of the spin operator). In the presence of the spin-orbit coupling, a non-vanishing source term could exist in the continuity equation of this conventional spin current \cite{culcer2004semiclassical,culcer2006geometrical,shi2006proper,Murakami2006conserved,Sugimoto2006PRB}, which is given by
\begin{equation}
    \frac{\partial s}{\partial t}+\nabla\cdot {\bf J}^{s}=\boldsymbol{\tau}.
\end{equation}
The source term is a torque corresponding to spin precession during the carrier propagation. Its operator is given by the time variation rate of the spin operator $\hat{\tau}=-i/\hbar[\hat{s},\hat{H}]$, where $\hat{H}$ is the Hamiltonian of the spin carrier. To exploit the continuity equation, we perform multipole expansion of the spin current ${\bf J}^{s}$ and the spin torque $\tau$ as \cite{xiao2021conserved}
\begin{eqnarray}
    &{\bf J}^{s} &={\bf J}^{{s},0}-\nabla\cdot {\bf D}^{\bf J}-\cdots,
    \label{expansion-1}\\
    &\tau &= -\nabla\cdot {\bf D}^{\tau}+\partial_{\alpha}\partial_{\beta}Q^{\tau}_{\alpha\beta}-\cdots,
    \label{expansion-2}
\end{eqnarray}
where ${\bf J}^{s,0}$ and ${\bf D}^{\bf J}$ represent the monopole and dipole densities of the conventional spin current, ${\bf D}^{\tau}$ and $Q^{\tau}_{ab}$ denote the dipole and quadrupole densities of the spin torque. Note that the monopole term of spin torque is always vanishing at the steady state, because the torque is a time derivative of a local physical quantity \cite{Sugimoto2006PRB,xiao2021conserved}. Hence, up to the second order of spatial gradient, the continuity equation can be cast into
\begin{equation}
        \frac{\partial s}{\partial t}+\partial_{\alpha}[ { J}^{s,0}_{\alpha}+ D_{\alpha}^{\tau}-\partial_{\alpha}(D^{J_{\alpha}}_{\beta}+Q^{\tau}_{\alpha\beta})]=0.
\end{equation}
From this sourceless continuity equation, one can define a conserved spin current \cite{xiao2021conserved}
\begin{equation}
    \mathcal{ J}^{s}_{\alpha}= {J}^{{s},0}_{\alpha}+ D_a^{\tau}-\partial_{\beta}(D^{J_{\alpha}}_{\beta}+Q^{\tau}_{\alpha\beta})+\cdots, \label{conserved current}
\end{equation}
It incorporates contributions from both the conventional spin current and spin torque. The detailed band expressions of the involved dipole and quadrupole densities can be found in Ref. \cite{Xiao2021conserved-arxiv,xiao2021conserved}, where a systematic multipole expansion method in the wavepacket theory is developed for evaluating all the terms in Eqs. (\ref{expansion-1}) and (\ref{expansion-2}).

\begin{table*} \label{tab:km}
    \centering
    \caption{Geometric quantities related to the superconducting mixed Berry curvature $\Omega_{\mathbf{km}}$}
   \begin{ruledtabular}
    \
    \begin{tabular}{cc}
      electronic Berry connection with respect to $\bf m$& $\mathfrak{A}^b_n({\bf k})=\langle\phi_{n{\bf k}}|i\nabla_{\bf m}\phi_{n{\bf k}}\rangle$\\
      electronic mixed Berry curvature&  $\Omega^b_{ k_{\alpha} m_{\beta}}({\bf k})=\partial_{k_{\alpha}}\mathfrak{A}^b_{n,\beta}({\bf k})-\partial_{ m_{\beta}}\mathcal{A}^b_{n,\alpha}({\bf k})$\\
      gauge-invariant phase derivative with respect to $\bf m$& $D_{\bf m}\varphi=\nabla_{\bf m}\varphi-\mathfrak{A}^b_n({\bf k})-\mathfrak{A}^b_n({-\bf k})$\\
      superconducting Berry connection with respect to $\bf m$&  $\mathfrak{A}_{\bf m}({\bf k})=-\frac{\rho}{2}\nabla_{\bf m}\varphi + |u|^2\mathfrak{A}_{n}^b(\mathbf{k})-|v|^2\mathfrak{A}_{n}^b(-\mathbf{k})$\\
      superconducting mixed Berry curvature & $\Omega_{k_{\alpha} m_{\beta}}({\bf k})= \frac{1}{2} ( D_{ k_{\alpha}}\varphi \partial_{ m_{\beta}} \rho   -\partial_{ k_{\alpha}} \rho  D_{m_{\beta}}\varphi) +|\mu|^2 \Omega^b_{ k_{\alpha} m_{\beta}} ({\bf k}) +|\nu|^2 \Omega^b_{k_{\alpha} m_{\beta}}(-{\bf k})$
    \end{tabular}
   \end{ruledtabular}
    
    \label{tab:km}
\end{table*}

This conserved spin current has several remarkable properties in spin thermoelectric transport, which are briefly described in order.

First, in uniform systems, the conserved current (\ref{conserved current}) reduces to 
\begin{equation}
    \mathcal{ J}^{s}_{\alpha}= {J}^{{s},0}_{\alpha}+ D_{\alpha}^{\tau}, 
\end{equation}
where the second term is the torque-dipole spin current \cite{shi2006proper}. The correct evaluation of $D_a^{\boldsymbol{\tau}}$ \textit{at equilibrium} was first presented in Ref. \cite{xiao2021conserved}. As a result, it turns out that in uniform systems
\begin{equation}
    \mathcal{ J}^{\bf s}_{\alpha}= 0 
\end{equation}
at equilibrium. This is a plausible property of conserved current, and is not shared by the conventional spin current.

Second, in nonuniform systems, the dipole of the conventional spin current ($D^{J_{\alpha}}_{\beta}$) and the quadrupole of the torque ($Q^{\tau}_{\alpha\beta}$) enter into the description of the conserved spin current, as shown in the gradient term of Eq. (\ref{conserved current}). As proven in \cite{xiao2021conserved}, it is these two quantities that certify microscopically
\begin{equation}
    \boldsymbol{\mathcal{J}}^{s}= \nabla\times {\bf M}^{s}
    \label{magnetization current}
\end{equation}
in nonuniform cases, i.e., the conserved spin current at equilibrium is a circulating magnetization current, as required to be so by the sourceless continuity equation.  ${\bf M}^{s}$ can be deemed as the orbital magnetization of spin, in analogy to the orbital magnetization of charge in electromagnetism. The expression of ${\bf M}^{s}$ has been figured out in \cite{xiao2021conserved}. 

Third, as is known in electromagnetism, the orbital magnetization current density ${\bf J}= \nabla\times {\bf M}$ is a circulating current, which cannot flow out of the sample as a result of the Stokes’ theorem \cite{Cooper1997}. For transport driven by temperature gradient, there is always the issue of subtracting the magnetization current from the microscopic current density to obtain the so-called transport current density. It is the transport current that can flow out of the sample. A prominent example is the anomalous Nernst effect, where the magnetization current given by Berry curvature must be subtracted to get the correct Nernst transport \cite{xiao2006berryphase}.
Here, in parallel with the charge thermoelectric transport, the contribution from the orbital magnetization of spin [Eq. (\ref{magnetization current})] must be subtracted from the total current to obtain the transport spin current. By doing so, the basic linear transport relations, including the Einstein relation, the Onsager relation \cite{xiao2020unified}, and the Mott relation, can all be established for spin transport \cite{xiao2021conserved}.

\section{Superconducting Berry quantities in the case of $\mathcal{T}$-invariant electronic Hamiltonian}\label{app:B}
If the single-electron Hamiltonian possesses $\mathcal{T}$ symmetry, then the electronic Berry curvature satisfies ${\bf \Omega}^{b}_{n}({\bf k})=-{\bf \Omega}^{b}_{n}(-{\bf k})$, and one can always choose a gauge such that ${\bf \mathcal{A}}_{n}^{b}({\bf k})={\bf \mathcal{A}}_{n}^{b}(-{\bf k})$. Thus, the superconducting Berry connection and Berry curvature reduce to
\begin{equation}
    \mathbf{\mathcal{A}}({\bf k})
=-\frac{1}{2}\rho \nabla_{{\bf k}}\varphi +  {\bf \mathcal{A}}_{n}^{b}({\bf k})
\end{equation}
and
\begin{eqnarray}
\mathbf{\Omega}_{\bf k}
=\frac{1}{2}\nabla_{\bf k}\varphi \times \nabla_{{\bf k}}\rho + {\bf \Omega}^{b}_{n}({\bf k}),
\end{eqnarray}
respectively. In these two expressions, the pairing induced part and the single-electron band part are independent and additive. Moreover, electronic Berry quantities at $\bf -k$ do not appear in the superconducting Berry quantities at $\bf k$. These two features are sharply contrasted with the case of $\mathcal{T}$-breaking electronic Hamiltonian shown in the main text.


\section{Superconducting mixed Berry curvature $\Omega_{\mathbf{km}}$}\label{app:C}
The Superconducting mixed Berry curvature is given by
\begin{equation}
    \Omega_{k_\alpha m_\beta}({\bf k})=\partial_{k_{\alpha}}\mathbf{\mathfrak{A}}_{n,\beta}({\bf k})-\partial_{ m_{\beta}}\mathbf{\mathcal{A}}_{n,\alpha}({\bf k}),
\end{equation}
where $\mathbf{\mathcal{A}}$ and $\mathbf{\mathfrak{A}}$ are the $k$-space and mixed Berry connections, respectively. The expressions for related geometric quantities are listed in Table \ref{tab:km}.

\section{Tight-binding model for the numerical illustration}\label{app:tight-binding}
For the convenience of reading, the tight-binding Hamiltonian of the  ferromagnetic Rashba model with s-wave superconductivity is given here
\begin{eqnarray}
    H_{\bf k}&&=[-2t(\cos k_x+\cos k_y)-\mu]\tau_z \\
    +\alpha_R&&(\sin k_y\tau_z\sigma_x-\sin k_x\sigma_y)
    +V_z\tau_z\sigma_z+i\Delta\tau_y\sigma_y,\nonumber
\end{eqnarray}
where $t$ denotes the nearest hopping energy, $\mu$ represents the chemical potential, $\alpha_R$ and $V_z$ are spin splitting energies of the Rashba-SOC and Zeeman field, and $\Delta$ stands for the $s$-wave pairing potential. Here, $\sigma$ and $\tau$ are Pauli matrices acting on spin space and  particle-hole space, respectively. 

\end{document}